%% file: ms.tex
\documentclass[preprint,12pt]{aastex}
\newcommand{\HETE}{{\it HETE}}

\newcommand{\WHnote}{in Gamma-Ray Burst and Afterglow Astronomy 2001,
AIP Conf. Proc 662, ed. G. R. Ricker \& R. K. Vanderspek (New York: AIP)}
\newcommand{\wxmonly}{{$^{b}$}}
\newcommand{\ergcmsqs}{{erg cm$^{-2} s^{-1}$}}
\newcommand{\ergcmsq}{{erg cm$^{-2}$}}
\newcommand{\epeak}{{$E_{peak}$~}}
\newcommand{\chisq}{{$\chi ^{2}$~}}
\newcommand{\etal}{et al.}
\newcommand{\gcn}{GCN}
\newcommand{\tfifty}{{$T_{50}$}}
\newcommand{\tninety}{{$T_{90}$}}
\newcommand{\degree}{$^{\circ}$}
\newlength{\GBCdigit}
\newcommand{\GBC}{\hspace*{\GBCdigit}}
\newlength{\GBCminus}
\newcommand{\hr}{\hspace{0.3em}}
\newcommand{\hl}{\hspace{-0.3em}}
\newcommand{\hL}{\hspace{-0.5em}}
\shorttitle{HETE-2 Observation of GRB030329}
\shortauthors{Vanderspek \etal}

\received{2004 January 12}
\begin{document}

\title{HETE Observations of
the Gamma-Ray Burst GRB030329: \\
Evidence for an Underlying Soft X-ray Component}

\author{
R.~Vanderspek,\altaffilmark{1}
T.~Sakamoto,\altaffilmark{2,3,4}
C.~Barraud,\altaffilmark{5}
T.~Tamagawa,\altaffilmark{3}
C.~Graziani,\altaffilmark{6}
M.~Suzuki,\altaffilmark{2}
Y.~Shirasaki,\altaffilmark{3,7}
G.~Prigozhin,\altaffilmark{1}
J.~Villasenor,\altaffilmark{1}
J.~G.~Jernigan,\altaffilmark{8}
G.~B.~Crew,\altaffilmark{1}
J.-L.~Atteia,\altaffilmark{5}
K.~Hurley,\altaffilmark{8}
N.~Kawai,\altaffilmark{2,3}
D.~Q.~Lamb,\altaffilmark{6}
G.~R.~Ricker,\altaffilmark{1}
S.~E.~Woosley,\altaffilmark{9}
N.~Butler,\altaffilmark{1}
J.~P.~Doty,\altaffilmark{1}
A.~Dullighan,\altaffilmark{1}
T.~Q.~Donaghy,\altaffilmark{6}
E.~E.~Fenimore,\altaffilmark{4}
M.~Galassi,\altaffilmark{4}
M.~Matsuoka,\altaffilmark{10}
K.~Takagishi,\altaffilmark{11}
K.~Torii,\altaffilmark{3,12}
A.~Yoshida,\altaffilmark{3,13}
M.~Boer,\altaffilmark{5}
J.-P.~Dezalay,\altaffilmark{14}
J.-F.~Olive,\altaffilmark{14}
J.~Braga,\altaffilmark{15}
R.~Manchanda,\altaffilmark{16}
and
G.~Pizzichini\altaffilmark{17}
}

\altaffiltext{1}{Center for Space Research, Massachusetts Institute of
Technology, 70 Vassar Street, Cambridge, MA, 02139.}

\altaffiltext{2}{Department of Physics, Tokyo Institute of Technology, 
2-12-1 Ookayama, Meguro-ku, Tokyo 152-8551, Japan.}

\altaffiltext{3}{RIKEN (Institute of Physical and Chemical Research),
2-1 Hirosawa, Wako, Saitama 351-0198, Japan.}

\altaffiltext{4}{Los Alamos National Laboratory, P.O. Box 1663, Los 
Alamos, NM, 87545.}

\altaffiltext{5}{Laboratoire d'Astrophysique, Observatoire
Midi-Pyr\'{e}n\'{e}es, 14 Ave. E. Belin, 31400 Toulouse, France.}

\altaffiltext{6}{Department of Astronomy and Astrophysics, University
of Chicago, 5640 South Ellis Avenue, Chicago, IL 60637.}

\altaffiltext{7}{National Astronomical Observatory, Osawa 2-21-1,
Mitaka,  Tokyo 181-8588 Japan.}

\altaffiltext{8}{University of California at Berkeley,
Space Sciences Laboratory, Berkeley, CA, 94720-7450.}

\altaffiltext{9}{Department of Astronomy and Astrophysics, University 
of California at Santa Cruz, 477 Clark Kerr Hall, Santa Cruz, CA
95064.}

\altaffiltext{10}{Tsukuba Space Center, National Space Development
Agency of Japan, Tsukuba, Ibaraki, 305-8505, Japan.}

\altaffiltext{11}{Faculty of engineering, Miyazaki University, Gakuen
Kibanadai Nishi, Miyazaki 889-2192, Japan.}

\altaffiltext{12}{Department of Earth and Space Science,
Graduate School of Science, Osaka University,
1-1 Machikaneyama-cho, Toyonaka, Osaka 560-0043, Japan.}

\altaffiltext{13}{Department of Physics, Aoyama Gakuin University,
Chitosedai 6-16-1 Setagaya-ku, Tokyo 157-8572, Japan.}

\altaffiltext{14}{Centre d'Etude Spatiale des Rayonnements,
Observatoire Midi-Pyr\'{e}n\'{e}es,
9 Ave. de Colonel Roche, 31028 Toulouse Cedex 4, France.}

\altaffiltext{15}{Instituto Nacional de Pesquisas Espaciais, Avenida
Dos Astronautas 1758, S\~ao Jos\'e dos Campos 12227-010, Brazil.}

\altaffiltext{16}{Department of Astronomy and Astrophysics, Tata 
Institute of Fundamental Research, Homi Bhabha Road, Mumbai, 400 005, 
India.}

\altaffiltext{17}{Consiglio Nazionale delle Ricerche,
IASF, Sezione di Bologna, via Piero Gobetti 101, 40129 Bologna, Italy.}

\begin{abstract}
An exceptionally intense gamma-ray burst, GRB030329, was
detected and localized by the instruments on board
the High Energy Transient Explorer satellite (\HETE) at 11:37:14 UT
on 29 March 2003. 
The burst consisted of two $\sim$10s pulses of roughly equal brightness
and an X-ray tail lasting $>$100s.
The energy fluence in the 30--400 keV energy band was
$S_\gamma = 1.2 \times 10^{-4}$ \ergcmsq, making GRB030329
one of the brightest GRBs ever detected.
Communication of a 2\arcmin~error box 73 minutes
after the burst allowed the rapid detection of a counterpart
in the optical, x-ray, radio and the ensuing discovery of a
supernova with most unusual characteristics.
Analyses of the burst lightcurves reveal the presence of a
distinct, bright, soft X-ray
component underlying the main GRB:
the 2--10 keV fluence of this component is $\sim$$7 \times 10^{-6}$ \ergcmsq.
The main pulses of GRB030329 were preceded 
by two soft, faint, non-thermal bumps.
We present details of the \HETE~observations of GRB030329.
\end{abstract}


\section{Introduction}

On March 29, 2003 the High Energy Transient Explorer
(\HETE; \citet{WH2001-HETE})
detected one of the brightest
gamma-ray bursts ever recorded, GRB030329. The intense flux
of this burst and the special characteristics of the HETE
mission allowed the 2\arcmin~localization of the burst 
to be communicated worldwide 72 minutes after
the onset of the burst. This led, in turn, to the rapid
discovery of optical, x-ray, and radio counterparts to the
burst, and, about one week later, to the discovery of one of
the most energetic, unusual supernovae ever observed, SN
2003dh \citep{stanek2003,hjorth03}.
This watershed event showed conclusively that at
least some long GRBs occur in conjunction with supernovae.  
Such an association was predicted by the collapsar model 
\citep{Woo93,Mac99}.

The instruments on \HETE~allow the study of prompt
radiation down to energies as low as 2 keV. 
The combination of high flux and low energy sensitivity has
allowed us to carry out very detailed analyses of the 
prompt emission of GRB030329.
These analyses have revealed soft features in the lightcurve:
faint, non-thermal, soft X-ray ``bumps'' before the
main GRB and a bright, soft X-ray component underlying
the main GRB.
In this paper, we describe the HETE observations of the prompt
emission of GRB030329.

\section{Discovery}

GRB030329 was detected by the \HETE~satellite 
\citep{WH2001-HETE}
as trigger H2652
at 11:37:14.7 UT (41834.7 SOD) on March 29, 2003
\citep{gcn1997}.
GRB030329 was seen as a very bright double-pulsed burst with
a duration of over 100s in all three HETE instruments:
Fregate: 7--400 keV \citep{WH2001-FREGATE};
Wide-field X-ray Monitor (WXM): 2--25 keV \citep{shirasaki03};
Soft X-ray Camera (SXC): 2--10 keV \citep{WH2001-SXC-jsv}.
GRB030329 had a 30--400 keV fluence of
$S_\gamma = 1.18\times10^{-4}$ \ergcmsq, making it one of the brightest GRBs
detected by \HETE~ to date
and placing it among the top 1\% of the brightest GRBs
ever detected.
The burst duration was
$t_{90} = 22.9$ s at high energies (30--400 keV), increasing to
$t_{90} \approx 38$ s at low energies (2--10 keV); a soft X-ray
tail is seen extending $>$100 seconds after burst onset. 

The burst was localized 
by the SXC
to a 2\arcmin~radius error circle (90\%)
centered at $\alpha$ = 10$^h$ 44$^m$ 49$^s$, 
$\delta$ = +21\degree~28\arcmin~44\arcsec~(J2000):
this position was distributed to the GRB Coordinates Distribution Network
(\gcn) 73 minutes after the
burst.  
\citet{gcn1986} began optical observations of the SXC error
region less than two minutes after the distribution of the GCN Notice,
\citet{gcn1985} fifteen minutes later; 
both observations
revealed a bright ($R<13$) optical transient (OT) just outside
the SXC error circle.
The bright OT was monitored by many observers over the following
days \citep{gcn1989,gcn1995,gcn1999,gcn2000,gcn2001,gcn2002,price03}.
\citet{greiner2003} reported a redshift of 0.1685, making GRB030329
the closest GRB with a measured redshift after GRB980425 
\citep{lidman98,tinney98} and
GRB031203 \citep{prochaska03,watson04}.
The emergence of the signature of a type Ic supernova from the
spectrum of the optical counterpart \citep{matheson2003,stanek2003,hjorth03}
solidifies the association 
of some GRBs with core-collapse supernovae.

Prompt followup observations revealed the enormous
brightness of GRB030329 at other wavelengths.
An RXTE observation of the X-ray afterglow starting less than five
hours after the burst revealed the source to have a 2--10 keV flux
of 1.4$\times 10^{-10}$ \ergcmsqs, one of the brightest GRB X-ray
afterglows ever measured \citep{gcn1996}.  
VLA observations at 8.46 GHz fourteen hours after the burst
\citep{gcn2014}
revealed a 3.5 mJy source, the brightest radio afterglow yet
measured.
The brightness of the radio afterglow was confirmed in the days following
the burst at a wide range of radio wavelengths
\citep{gcn2043,gcn2073,gcn2088,gcn2089,gcn2161}.
GRB030329 was also detected as
a Sudden Ionospheric Disturbance \citep{gcn2176},
as
the incoming X-ray flux was sufficiently strong
to create a global surge in the electron density on the illuminated side
of the Earth.

\section{Observations} \label{observations}

GRB030329 was detected by Fregate as an 8.3$\sigma$ excess in the 
7--80 keV band on the 1.3s timescale trigger.
This excess was due to a soft, relatively weak precursor:  thirteen
seconds later, the burst was as bright as any seen during the
\HETE~mission.
Real-time reports of the burst brightness via the \HETE~VHF
system 
\citep{WH2001-BAS-gbc}
showed the significance of the burst in the WXM exceeding
the 7 bits allotted in the telemetry stream at S/N=127.
At the same time, the image S/N reported from the WXM was
60, twice that of the previous brightest burst (GRB020813).

Ground analyses of the downlinked data starting $\sim$40 minutes
after the trigger confirmed the magnitude of the burst.
The burst was detected at a S/N of over 300 in both Fregate
and the WXM, and over 100 in the SXC:  this burst is the
brightest \HETE~has detected in the 7--80 keV band.
Unfortunately, the burst was {\em so} bright and {\em so} long that the Fregate
flight software decided the burst was due to a increase in the overall
background rate and ``invalidated'' the burst in flight.
As a result, the initial real-time \gcn~messages indicated that H2652
was {\em not} a real GRB.
More important, however, is the fact
that time-tagged photons are not downlinked from the Fregate
instrument for bursts that are
invalidated in flight\footnote[1]{This same scenario occurred 24
hours earlier on the bright burst GRB030328
(H2650), the only two times in the course of the \HETE~ mission
that real GRBs have been invalidated in flight.  The flight software
has since been modified to prevent bright bursts from being
invalidated.}.
The temporal and spectral analyses of Fregate data were 
therefore done with survey data:  lightcurves in three broad energy bands
(band A:  7--40 keV; band B:  7--80 keV; band C:  30-400 keV)
at 80 ms time resolution and 128-channel spectral data products
with a time resolution of 5.2 seconds.

\subsection{Localization} \label{localization}

GRB030329 was very bright in X-rays:  the burst was
detected at a S/N of 350 in the WXM.
However, the large incident
angle of the burst was such that only one of the two X
detectors and one of the two Y detectors of the WXM were
illuminated.
Unfortunately, the Y detector illuminated by the burst was
YB, which was lost due to a micrometeorite impact in
January 2003;
as a result, while the WXM could be used to determine the X position 
very precisely, the Y position could only be crudely determined,
based on the relative illumination of the six wires in the X detectors.
Fortunately,
despite the large incident angle, GRB030329 was well detected
in both the SXC X and Y detectors.

The real-time localization of GRB030329 calculated in
flight reflected these facts:
the image S/N calculated on board (which reflects the strength
of the peak in the cross-correlation map) was 60 in X, but less
than 2 in the Y direction.
As the SXC flight imaging software, which looks for significant
cross-correlation peaks near the WXM localization, requires
image S/N values greater than 3 in both WXM axes to proceed,
no real-time SXC localization was available.

\begin{figure}
\epsscale{0.70}
\plotone{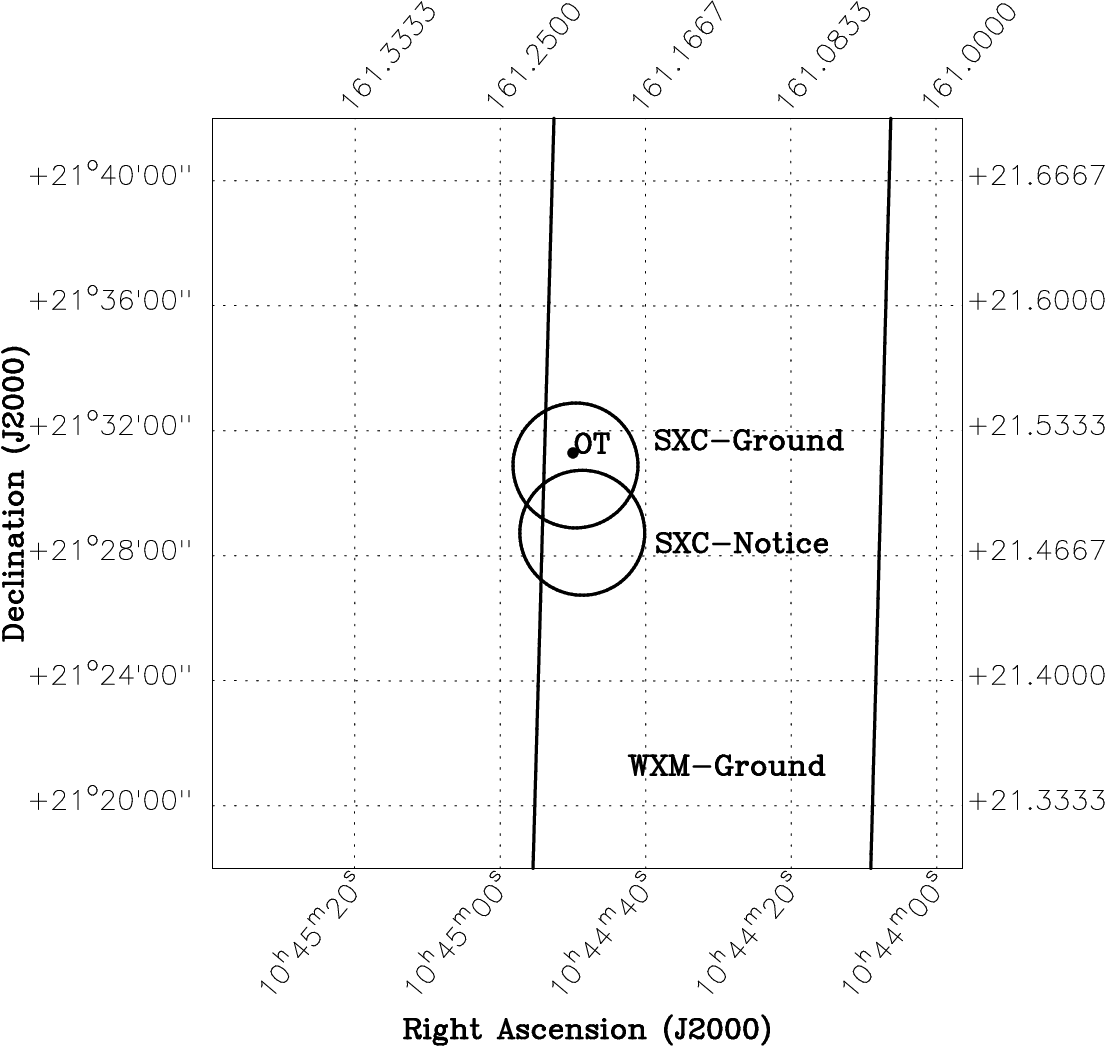}
\figcaption{The localization history of GRB030329.  Shown are a
portion of the 12\arcmin$\times$2.25\degree~WXM error region,
the first SXC position (SXC-Notice), the SXC position
corrected for the large systematic error at this
burst's incident angle (SXC-Ground), and the location of the OT.
The SXC error regions are 2\arcmin~ in radius (90\%).
\label{fig:skymap}}
\end{figure}

Ground analyses of the WXM and SXC data began after the full
data set was downlinked, at 12:10 UT (33 minutes after the burst).  
It quickly became clear
that only the X detector was illuminated, limiting the WXM
localization to a box measuring 12\arcmin$\times$2.25\degree.
Fortunately, the burst
was so bright in the SXC that it was possible to calculate a position
without reference to the WXM position, and the SXC position
was in the WXM error box.  
The SXC localization was distributed by GCN Notice at 12:50 UT
(73 minutes after the burst) as a 2\arcmin~circle centered on
$\alpha$ = 10$^h$ 44$^m$ 49$^s$, 
$\delta$ = +21\degree~28\arcmin~44\arcsec~ (J2000).  
Continued analyses of the SXC data after the distribution of
the first GCN Notice revealed that a correction to the localization
to account for systematic errors at large incident angles
had not been applied, and that the procedure to distribute
such localizations with a larger quoted error circles
had not been followed.
A revised error region that compensates for the systematic
error was distributed in \citet{gcn1997}: 
it is a 2\arcmin~ circle centered on
$\alpha$ = 10$^h$ 44$^m$ 50$^s$, 
$\delta$ = +21\degree~30\arcmin~54\arcsec~(J2000).
The skymap of GRB030329,
showing the WXM and SXC localizations and the OT, is shown in Figure 
\ref{fig:skymap}.

GRB030329 was observed by every spacecraft in the Interplanetary
Network (Ulysses, Konus-Wind, Mars Odyssey (HEND and GRS),
RHESSI, and INTEGRAL (SPI-ACS))
and the resulting IPN localization is fully consistent with,
but does not constrain,
the WXM and SXC localizations (K. Hurley 2003, private communication).

\subsection{Temporal Properties} \label{time_history}

\begin{figure}
\includegraphics[width=3.4in]{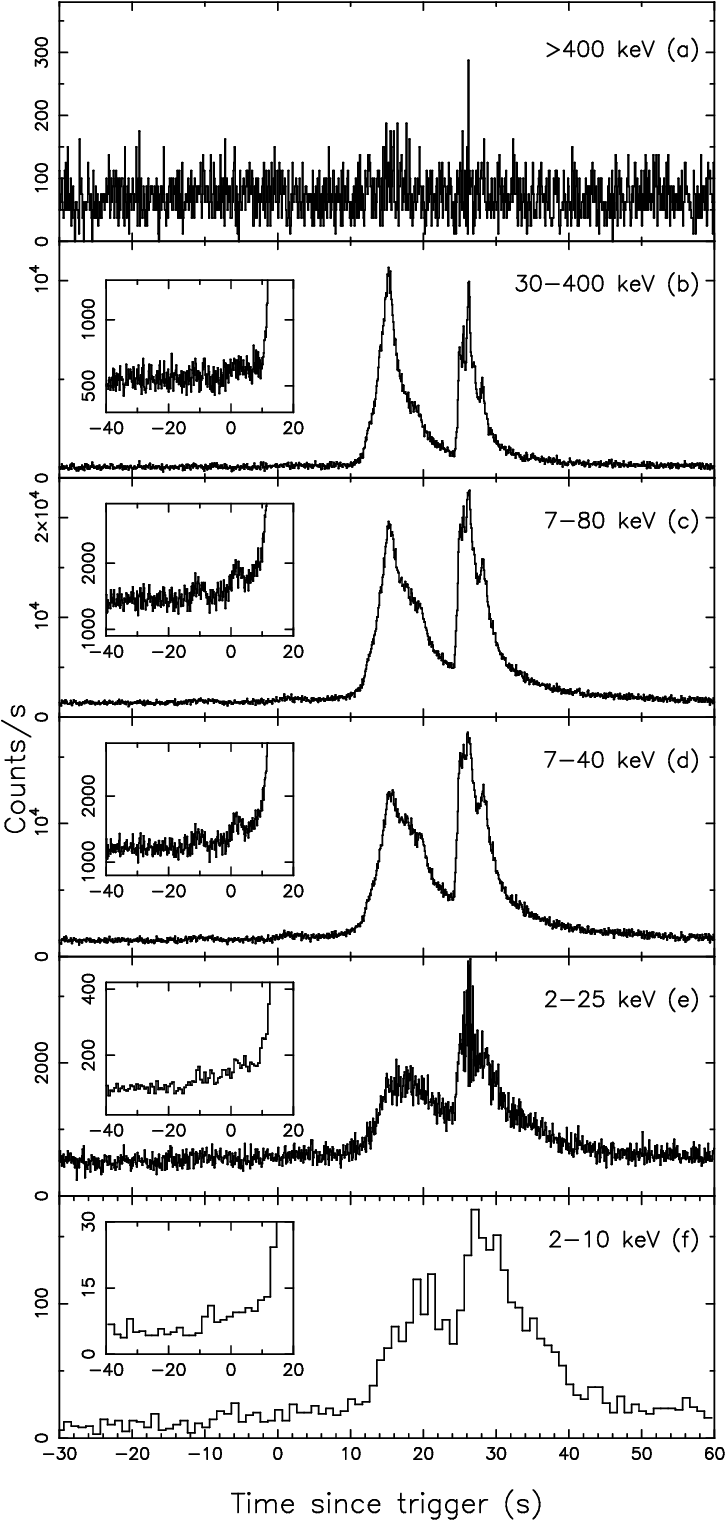}
\figcaption{Time histories of GRB030329.
Shown are the lightcurves in 
(a) Fregate band D,
(b) Fregate band C,
(c) Fregate band B,
(d) Fregate band A,
(e) the WXM, and 
(f) the SXC.
The inset figures show more detail for the burst precursors in each band.
The time resolution of the Fregate plots is 80 ms (160 ms in the insets);
for the WXM, it is 80 ms (640 ms); for the SXC, it is 1 s (2 s).  
The broad time bands may hide significant spectral evolution at the
start of the second pulse:  the short
spike in band D at t=26.2s indicates that the burst was 
harder for a brief period at the start of the second pulse.
\label{fig:LCs}
}
\end{figure}

As shown in Figure \ref{fig:LCs},
the time profile of GRB030329 is dominated by two strong pulses
separated by 11s and an extended soft tail that continues
$>$60 seconds after the trigger.
The Fregate band D ($E > 400$ keV) lightcurve shows a $\sim$6$\sigma$
enhancement over background during the first pulse and
a single 80 ms bin ``spike'' at t=26.2s, coincident with the
peak of the second pulse.
The burst duration was
$t_{90} = 22.9$ s at high energies (30--400 keV), increasing to
$t_{90} \approx 38$ s at low energies (2--10 keV):
Table \ref{tbl:temporal} gives the \tfifty~and \tninety~ durations for
various SXC, WXM and FREGATE energy bands
\citep{t90ref}.  
Both duration measures decrease with increasing energy:
this behavior is typical of many long-duration GRBs \citep{fenimore95}. 

There is clear spectral evolution
during the burst, as the relative brightness of the first
pulse to the second pulse is larger in Fregate band C 
than in band A.
There is also distinctly different spectral behavior in 
the two pulses.
In Figure \ref{fig:C_SXC}, we overlay the SXC (2--10 keV)
and Fregate band C lightcurves:
while the onset of the second pulse
is essentially simultaneous in all energy bands, there is a
delay in the peak of the first pulse of $\sim$5 seconds between
photons with E$>$25 keV and those with E$<$10 keV.
The nature of this evolution is discussed in more detail in
Section \ref{sec:specdisc}.

The inset figures of Figure \ref{fig:LCs} reveal two small
precursor bumps, seen primarily in Fregate bands A and B.
(The second bump in band B is in fact the trigger for
this burst in Fregate).
While the bumps are not as significant in the WXM and SXC
lightcurves,
localization analyses of both WXM and SXC data show the
photons came from the same celestial coordinates as the
photons from the two main pulses:  this result holds for
photons detected as early as $t=-13s$.
Integrating a pair of gaussians fit to the two bumps
shows that the total counts in the two bumps is roughly
1\% of the total in the two main pulses in both bands
A and B.
In each bump, the 7--30 keV counts fluence is $\sim$90\% of
the 7--80 keV counts fluence, indicating that the bulk of
the precursor emission is $<$30 keV;
the fact that the second bump is slightly more prominent than the first one
in band C (see the inset of \ref{fig:LCs}b)
may indicate that the second bump is somewhat harder than the first.
Also visible in the inset figures in Figure \ref{fig:LCs}
is a general rise in the count rates at lower energies
in the period leading up to the first major pulse.
The burst begins in the 30--400 keV band at $t \approx -4$ seconds, while
2--25 keV photons are first detected at $t \approx -13$ seconds.

GRB030329 also has a soft X-ray tail which continues $>$100s after
the trigger.  
T. Tamagawa et al (2004, private communication) have shown that 
an extrapolation of the 2--10 keV photon energy flux of the afterglow 
\citep{tiengo2003,gcn2052}
back to t=100s is consistent with the 2--10 keV photon energy 
flux observed in the X-ray tail at that time;
the spectral index of the tail ($\sim$-2.1) is also consistent 
with that cited in \cite{tiengo2003}.
The soft X-ray tail of GRB 030329 
will be discussed in more detail in \citet{atteia04}.

\subsection{Spectrum} \label{spectrum}

Because the burst was invalidated in flight, there are no
time-tagged photon data from the Fregate instrument available.
However, survey spectral data, which consists of 128 spectral
bins every 5.2 seconds, can be used to calculate the burst spectral
parameters.
The spectrum of the entire burst
can be well fit by a Band function from 2--400 keV, with 
parameters 
$\alpha=-1.32 \pm 0.02$,
$\beta=-2.44 \pm 0.08$,
and \epeak = $70.2 \pm 2.3$ (errors are 90\%);
however, in time-resolved spectroscopy, these parameters
vary significantly during the course of the burst.
The burst fluence was 
$S_\gamma = 1.2\times 10^{-4}$ \ergcmsq~in the
30--400 keV band, 
$S_X = 5.5\times 10^{-5}$ \ergcmsq~in the 2--30 keV band:
the hardness ratio $S_X / S_\gamma$ is 0.56, which qualifies
GRB030329 as an X-ray rich burst \citep{heise2001}.
At the measured redshift of 0.1685, the total isotropic
energy released was $(1.86 \pm 0.08) \times 10^{52}$~erg and
$E_{peak}$ in the source frame is $82.0 \pm 2.6$~keV,
consistent with the relationship described in \citet{amati02}.
The time-integrated emission properties of GRB030329 in various
energy bands are given in Table 2.

The results of time-resolved spectral analyses 
are given in Table \ref{table:specparams} and plotted
in Figure \ref{fig:sp_figure}.
The Fregate and WXM data were fit jointly, to a Band function
when possible and a power-law model when a significant value 
of Band $\beta$ could not be calculated.
The data show a strong softening of the burst spectrum
during the course of the first pulse, followed by
a mild re-hardening at the onset of the second pulse
and a subsequent softening.

Power-law fits to the WXM precursor data
reveal that the photon indices of the two
bumps are similar, $-2.0 \pm 0.4$.
A joint fit of a power-law model to the WXM and Fregate data
for both bumps together (from t=-13s to t=+3s)
shows a photon index
of $-2.11 \pm 0.09$, a 2--25 keV fluence of
1.2$\times 10^{-6}$ \ergcmsq, and a 30--400 keV fluence of
9.3$\times 10^{-7}$ \ergcmsq.
(Fits of the precursor 
data to a cutoff power law (Comptonization) model and a Band 
function were also performed, but neither resulted in a
significant improvement in $\chi$$^2$).
The power-law fit is quite good
(reduced \chisq of 0.7):
the precursor spectra are very clearly
non-thermal.

\section{Soft X-Ray Emission and Precursor Activity} \label{discussion}

\subsection{Spectral Lag or Soft X-ray Bump?}
\label{sec:specdisc}
\begin{figure}
\includegraphics[angle=270,scale=0.4]{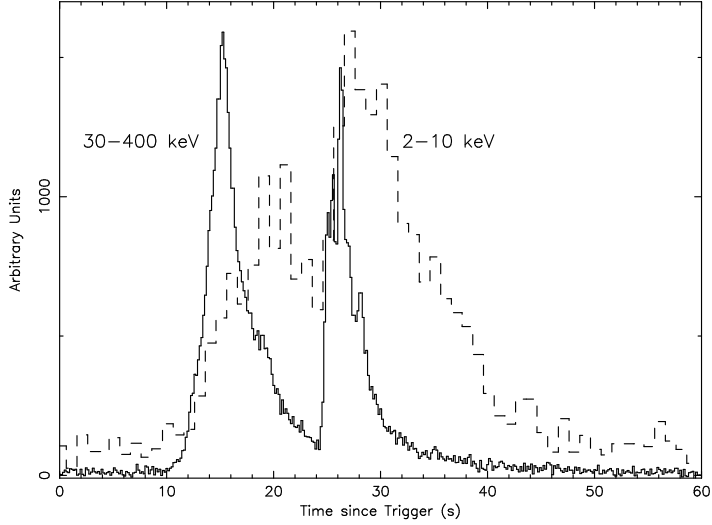}
\figcaption{An overlay of the Fregate band C (30--400 keV; solid line) and
SXC (2--10 keV; dashed line) lightcurves for GRB030329 
show the distinct nature
of the two pulses of GRB030329.
There is a delay of $\sim$5 seconds between the peak of
the 30--400 keV and 2--10 keV emission in the first pulse,
while the two are essentially simultaneous in the second
pulse. (The SXC lightcurve has been scaled for clarity).
\label{fig:C_SXC}}
\end{figure}

\begin{figure}
\includegraphics[angle=270,scale=0.4]{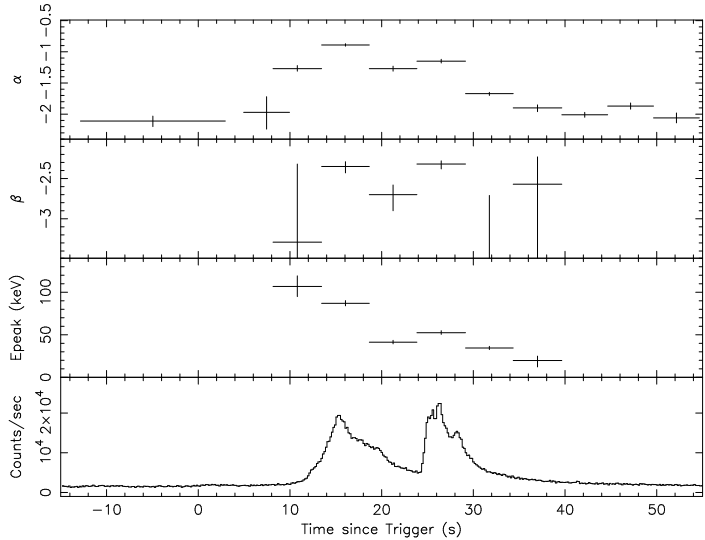}
\figcaption{Variation of the spectral parameters of GRB030329 with time.
Shown are the Band $\alpha$, $\beta$, and \epeak parameters with
the band B lightcurve for reference; for
regions where Band $\beta$ cannot be calculated, $\alpha$ is
the index in a fit of a power law to the counts spectrum.
\label{fig:sp_figure}}
\end{figure}

At first glance, the broadband lightcurves shown in Figure \ref{fig:LCs}
seem to indicate that there is a strong
hard-to-soft evolution in the burst emission in the first pulse.
However, closer evaluation of the burst lightcurves suggests that
this ``evolution'' is more likely due to
a relatively bright, soft X-ray bump of $\sim$20 seconds duration
centered at t $\approx$ 20s after the burst trigger, as shown in the
overlay of band C and SXC emission in Figure \ref{fig:C_SXC}.

We examined the GRB030329 burst data for spectral lags of the
type discussed in \citet{band97}.
This analysis was made difficult because the Fregate 
broadband lightcurve bandpasses
do not match up well with the BATSE discriminator bandpasses
(20--50 keV, 50--100 keV, 100--300 keV) and because there are
no time-tagged photon data with which to reconstruct such lightcurves; 
however, linear combinations of Fregate bandpass lightcurves can
be used to shed some light on the spectral evolution of the first pulse
of the burst.
We calculate the difference of Fregate band B (7--80 keV) and 
Fregate band A (7--40 keV) to create a Fregate band ``B-A'' with
an effective bandpass of 40--80 keV; we also construct a
``C-(B-A)'' lightcurve, which has an effective energy bandpass
of 30--40 keV {\em plus} 80--400 keV.
A simple cross-correlation of the band B-A lightcurve with the
band C-(B-A) lightcurve (160 ms time resolution) shows a ``lag''
of 6 $\pm$ 32 milliseconds.
This value indicates that there is essentially no shift in the
temporal structure of the GRB030329 lightcurve with energy at
energies greater than 30 keV, and thus that the strong spectral
evolution in the first pulse must be due to photons with E$<$30 keV.

\begin{figure}
\includegraphics[angle=270,scale=0.4]{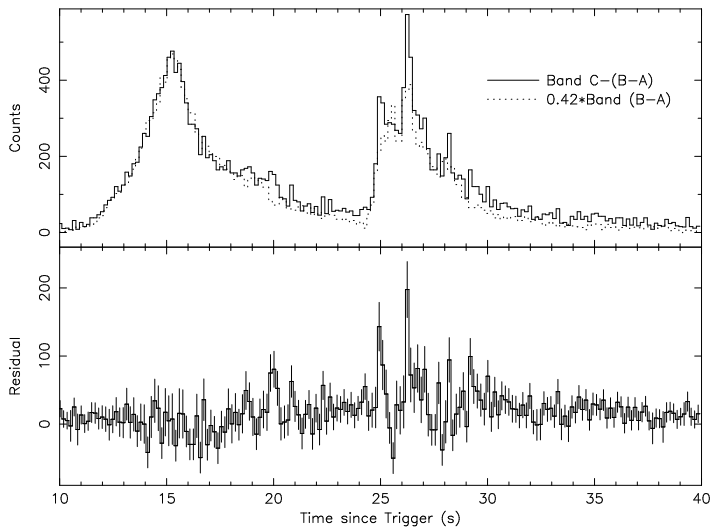}
\figcaption{
The 40--80 keV lightcurve of GRB030329,
calculated as Fregate band B - Fregate band A,
is scaled to the 30--400 keV minus the 40--80 keV lightcurve
using a one-second
interval centered on the peak of the first pulse (t=14.5 to t=15.5).
The 40--80 keV lightcurve is multiplied by the ratio of the counts
fluence in that time period (0.42) and compared to 
the band C-(B-A) lightcurve.
The two lightcurves are shown in the upper panel:
the solid line is band C-(B-A), the dotted line is 0.42*band(B-A).
The difference between the two (lower panel) reflects excesses in either the
30--40 keV or the 80--300 keV band with respect to the template
defined by the B-A lightcurve.
The difference is consistent with zero except for
a peak at t=20s (coincident with the peak of the soft X-ray
emission in the first pulse), 
two short spikes - - at t=25s (coincident with the onset of the
second pulse) and t=26.2s (coincident with the spike in 
band D) - - and a faint
extended tail starting at about t=28s (consistent with an extended
soft tail to the burst).
We conclude that the overall shape of the lightcurve 
of GRB030329 does not vary significantly from 30 to 400 keV, and
thus that the soft X-ray pulse seen in Figure \ref{fig:C_SXC}
could be due to a different emission mechanism than the main burst.
\label{fig:scaled}}
\end{figure}

As an alternate approach to understanding the temporal 
and spectral variability of the main pulses, we scale
the B-A lightcurve to the C-(B-A) lightcurve and examine
the differences between the two.
The scaling is done using 
only the data at the peak of the first pulse (between t=14.5s and 
t=15.5s after the trigger):  the resultant scale factor is 0.42.
In Figure \ref{fig:scaled} we show the band C-(B-A) lightcurve, the
scaled band B-A lightcurve, and the difference between the two:
the shapes of the lightcurves
match extremely well for the duration of the burst, even 
though the scaling was done using data only from the maximum
of the first pulse.
The difference reflects differences in the temporal
characteristics of the [30--40] + [80--400] keV band with respect to the
template 40--80 keV band and could be due to features in
either the 30--40 keV band or the 80--400 keV band.
The difference is consistent with
zero except for a few pulse-like features 
- - most notable are 
a peak at t=20s, consistent with the peak of the soft X-ray 
emission (Figure \ref{fig:C_SXC}), 
a peak at the start of the second pulse (t=25s), 
and a spike coincident with the band D spike (t=26s) 
- - and a faint tail starting at t=28s.
Interpretation of the details of this difference is not straightforward;
however, this analysis confirms that the extreme ``lag'' implied by the 
lightcurves in Figures \ref{fig:LCs} and \ref{fig:C_SXC}
is not seen at energies $>$30 keV.

The more likely explanation for the time profiles seen in
Figures \ref{fig:LCs} and \ref{fig:C_SXC}
is a bump of soft X-ray emission 
which begins at about the same time as the main
GRB and peaks at t $\approx$ 20s, five seconds after the peak of
the first pulse.  
Analysis of the 2--10 keV lightcurve show that
a gaussian with a FWHM of $\sim$12 seconds is a reasonable fit to
that pulse:  
we estimate the 2--10 keV fluence of this pulse to
be $\sim 7 \times 10^{-6}$ \ergcmsq~
(because the spectrum of GRB itself extends into the soft X-ray
band, it is not possible to clearly separate the 
first pulse into the sum of two distinct lightcurves).

Detailed spectral analyses of the soft X-ray pulse are made difficult
by the overlap with the first GRB pulse and the absence of 
high-time-resolution spectral data; 
thus, it is not possible to demonstrate the existence of the 
soft X-ray pulse by fitting the spectrum of GRB030329 to a two-component
spectral model.
The excellent fit of the burst spectrum to
a GRB model \citep{band93} in that time period 
indicates that the soft pulse likely has a non-thermal spectrum.
If the broadband lightcurve data are not considered, it would be 
easy to disregard this soft emission as part of the spectral evolution 
of the GRB; however,
the dissimilarity in the 2--10 keV and band C lightcurves in the first
pulse and the stark contrast in the burst emission profiles above 30 keV
vs. below 20 keV strongly suggest that the soft X-ray pulse is
indeed a distinct emission component.

The behavior of the soft X-rays of the second pulse is entirely different
from that of the first pulse.
The fast rise of the 2--10 keV count rate in the second pulse is 
coincident with the onset of the second pulse in the band C lightcurve, 
in contrast to
the delayed rise of the 2--10 keV component of the first pulse.
There is no evidence for a distinct soft X-ray component in the second 
pulse:
the similarity of the band C and 2--25 keV time profiles in the second
pulse indicates that the emission in these two bands likely come from
the same source, with any variations in relative brightness being
due to spectral parameters changing as a function of time
(see Figure \ref{fig:sp_figure}).

\subsection{Precursors}

The concept of a burst ``precursor'' has various definitions
when applied to GRBs.
The original definitions were ostensibly ``$\gamma$-rays seen
before the pulse that triggered the instrument'' or
``X-rays seen before the $\gamma$-rays''.
The term ``precursor'' has been applied to 
$\gamma$-ray pulses before the event that triggered BATSE
\citep{koshut95}, 
the slow onset
of soft X-ray emission before the main $\gamma$-ray pulse
(e.g., GRB900623; \citet{murakami91}),
as well as isolated
X-ray pulses before the GRB itself (e.g., GRB790307, \citet{laros84}).

\citet{koshut95} defined a precursor to be emission which precedes
the main burst by at least the duration of the main burst;  
using this restrictive definition and BATSE DISCLA data with
E $>$ 25 keV, they found that the distribution of ``precursor''
spectral parameters is not significantly different than the distribution
of GRB spectral parameters.
On the other hand,
\citet{grb980519} distinguish precursor emission from the burst
emission by {\em requiring} that the precursor emission be described
by an entirely different spectral model than the burst emission
or that the spectral parameters of the precursor be unprecedented
in describing GRBs in the $\gamma$-ray phase.
According to \citeauthor{grb980519}, then,
the thermal emission of GRB900126 \citep{murakami91}
is a true precursor, while the ``X-ray bumps before the GRB'' in
GRB980519 and GRB990704 \citep{feroci01} are not.
\citeauthor{koshut95} would not consider any of these examples
to be true precursors; however, these precursors are so soft that
they would not have been considered in their analysis.

The features in the low-energy lightcurves of GRB030329 at t $\approx$ -10 s
and t $\approx$ 2 s are very similar to the ``X-ray bumps before the
GRB'' in GRB980519.
In that burst, detected by BeppoSAX, three soft X-ray bumps
precede the main burst by 10--40 seconds; the bumps are
non-thermal and are well-fit by a power law of index $\sim$-2:
\citet{grb980519} consider these bumps to be part of the main burst
emission.
In fact, the spectral evolution of GRB980519 overall is quite
similar to that of GRB030329, as the power-law index is $\sim$-1
at the peak of the burst, then settles to values near -2 as the
burst fades.  
GRB980519 is significantly fainter, however:  the 25--300 keV 
fluence is $7 \times 10^{-6}$ \ergcmsq, roughly fifteen times
fainter than GRB030329.
Scaling from Figure 2 of \citet{grb980519}, the 2--10 keV
fluence of the ``X-ray bumps before the GRB'' of GRB980519 is
$\sim 5 \times 10^{-7}$ \ergcmsq, which, for a power law of
index -2, is comparable to the
2--25 keV fluence of $\sim10^{-6}$ \ergcmsq~for
GRB030329.
It is unclear whether the fact that these pre-GRB X-ray fluences 
are so similar for bursts that differ by $>10\times$ in overall
fluence is an
indication that an independent mechanism is responsible for
the creation of the pre-GRB X-rays; the detection
of more bursts of this kind are required to understand this better.
Several bursts in the HETE database have similar X-ray ``precursors'':
these will be discussed in a separate paper.

\section{Conclusions}

GRB030329 was seen as a two-peaked GRB with
$S_\gamma = 1.2\times 10^{-4}$ \ergcmsq, making it 
the brightest GRB detected by HETE and one of the 
brightest GRBs ever detected.  
The substantial X-ray fluence of GRB030329 ($5.5\times 10^{-5}$ \ergcmsq)
puts it in the class of X-ray Rich GRBs.
The overall burst can be fit will with a GRB model,
with $\alpha = -1.32 \pm 0.02$, $\beta = -2.44 \pm .08$,
and $E_{peak} = 70.2 \pm 2.3$ keV, but there is substantial
spectral variability during the burst.

Detailed analyses of the burst lightcurves reveal the presence
of a soft X-ray component distinct from the main GRB.
This component consists almost entirely of photons
with energies $<$25 keV, and it is quite bright:  the
2--10 keV fluence is $\sim 7 \times 10^{-6}$ \ergcmsq.
The burst lightcurves also show two soft, non-thermal X-ray
``precursor'' bumps; while such pre-gamma-ray peak X-ray 
emission has been detected in some GRBs, it is not a common feature
of GRB lightcurves.

The enormous brightness of the optical afterglow of GRB030329 has
allowed it to be used as a ``Rosetta Stone'' in understanding the
correlation of GRBs with Type Ic supernovae.
The analysis of the bright prompt X- and $\gamma$-ray emission
of GRB030329 has revealed a bright, soft X-ray component apparently
uncorrelated with the main body of the burst emission.
The obvious presence of this feature in a burst as bright as
GRB030329 leads to the question of whether this feature is present
in other GRBs and whether the brightness of the feature scales
with burst fluence.
This underlying soft pulse would not have been detected by the BATSE LADs,
as detectors sensitive to photons with E$<$20 keV are required
to detect these pulses;
however, at certain gain settings,
the BATSE SDs would have detected the soft emission of such an 
intense event \citep{preece00}.
Searches for such features in GRBs detected by HETE are underway; we 
urge that similar analyses of BeppoSAX, Ginga, and BATSE SD data be performed.
In this way, perhaps GRB030329 will also act as a ``Rosetta Stone'' in 
furthering our understanding
of the prompt X- and $\gamma$-ray emission in long GRBs.

\acknowledgments
\section*{Acknowledgments}

The authors acknowledge useful discussions with Enrico Ramirez-Ruiz,
Chryssa Kouveliotou, Jay Norris, Jon Hakkila, 
Marco Feroci, and Filippo Frontera.
The authors also thank the referee for a thorough reading of the
manuscript and several insightful comments.
The HETE mission is supported in the US by NASA contract NASW-4690; in
Japan, in part by the Ministry of Education, Culture, Sports, Science,
and Technology Grant-in-Aid 13440063; and in France, by CNES contract
793-01-8479.
KH is grateful for HETE support under Contract MIT-SC-R-293291, for Ulysses
support under JPL Contract 958056, and for IPN support under NASA grant
FDNAG5-11451.
SW is thankful for support under the NASA Theory Program under
grant NAG5-12036.
G. Pizzichini acknowledges support by the Italian Space Agency.

\clearpage

\begin{deluxetable}{lccc}
\tablecaption{Temporal Properties of GRB030329.
\label{tbl:temporal}}
\tablewidth{0pt}
\tablehead{
\colhead{Instrument} & \colhead{Energy}
& \colhead{\tfifty} & \colhead{\tninety} \\
& \colhead{(keV)} & \colhead{(s)} & \colhead{(s)}
}
\startdata
HETE SXC     & \GBC2--10\GBC    & 12.2  $\pm$ 0.3  &  34.7 $\pm$ 1.5  \\
&&& \\
HETE WXM     & \GBC2--25\GBC    & 12.8  $\pm$ 0.2  &  41.6 $\pm$ 0.8  \\
             & \GBC2--5\GBC\GBC & 14.5  $\pm$ 0.4  &  52.1 $\pm$ 1.2  \\
             & \GBC5--10\GBC    & 12.4  $\pm$ 0.3  &  39.8 $\pm$ 1.0  \\
             & 10--25\GBC       & 11.5  $\pm$ 0.2  &  36.9 $\pm$ 1.1  \\

&&& \\
HETE FREGATE & \GBC7--40\GBC    & 11.2 $\pm$ 0.06 & 29.1 $\pm$ 0.6    \\
             & 30--400          & 11.2 $\pm$ 0.05 & 22.8 $\pm$ 0.5    \\
\enddata
\vskip -18pt
\tablecomments{Errors are 1$\sigma$; the SXC energy range is approximate.}
\end{deluxetable}

\clearpage


\begin{deluxetable}{ccccc}
\tablecaption{Emission Properties of GRB030329.
\label{tbl:emission}}
\tablewidth{0pt}
\tablehead{
\colhead{Energy} &
\colhead{Photon Flux} &
\colhead{Energy Flux} & 
\colhead{Energy Fluence} \\
\colhead{(keV)} & 
\colhead{(ph cm$^{-2}$ s$^{-1}$)} & 
\colhead{($10^{-7}$~erg cm$^{-2}$ s$^{-1}$)} & 
\colhead{($10^{-5}$~erg cm$^{-2}$)}
}
\startdata
2--10    & $57.5  \pm  1.1$ &  $4.20  \pm 0.07$ &  $2.64  \pm 0.04$ \\
2--25    & $77.0  \pm  1.1$ &  $9.16 \pm 0.08$ &  $5.76  \pm 0.05$ \\
2--30    & $79.9  \pm  1.1$ &  $10.43 \pm 0.08$ &  $6.57  \pm 0.05$ \\
7--30    & $32.35 \pm 0.19$ &  $7.53  \pm 0.04$ &  $4.74  \pm 0.03$ \\
30--400  & $15.09 \pm 0.10$ &  \hL$18.78 \pm 0.25$ &  \hl$11.82 \pm 0.16$ \\
50--100  & \hr$5.39  \pm 0.05$ &  $5.98 \pm 0.06$ &  $3.76  \pm 0.04$ \\
\hL 100--300 & \hr$2.91  \pm 0.05$ &  $7.35 \pm 0.16$ &  $4.62  \pm 0.10$ 
\enddata
\vskip -18pt
\tablecomments{Errors are for 90\% confidence.  All values were
calculated from the time-averaged Band spectrum of the entire burst.}
\end{deluxetable}

\clearpage
\input{tab3.tex}
\clearpage

\end{document}

%% file: tab3.tex
\begin{deluxetable}{cccccc}
\tablecaption{Spectral Model Parameters for GRB030329.
\label{tbl:spectrum}}
\tablewidth{0pt}
\tablehead{
\colhead{Time since}
	& \colhead{}
	& \colhead{}
	& \colhead{E$_{peak}$}
	& \colhead{2--25 keV}
	& \colhead{30--400 keV} \\
\colhead{trigger (s)}
	& \colhead{$\alpha ^{a}$}
	& \colhead{$\beta$}
	& \colhead{(keV)}
	& \colhead{Flux}
	& \colhead{Flux}
}
\startdata
 -10.05 -- -5.05\wxmonly
	& $-2.16~_{-0.44}^{+0.38}$
	& ...
	& ...
	& \hspace{0.6em}$0.66~_{-0.17}^{+0.18}$
	& ... \\ 
\hspace{0.4em}-5.05 -- -0.05\wxmonly
	& $-1.96~_{-0.56}^{+0.49}$
	& ...
	& ...
	& \hspace{0.6em}$0.68~_{-0.20}^{+0.23}$
	& ... \\ 
\hspace{0.1em}-0.05 -- 4.95\wxmonly
	& $-2.01~_{-0.30}^{+0.31}$
	& ...
	& ...
	& \hspace{0.6em}$1.18~_{-0.20}^{+0.24}$
	& ... \\ 
\hspace{0.4em}4.95 -- 9.95\wxmonly
	& $-1.97~_{-0.27}^{+0.25}$
	& ...
	& ...
	& \hspace{0.6em}$1.37~_{-0.22}^{+0.23}$
	& ... \\ 
\hspace{-0.8em}-12.85 -- 2.95 
	& $-2.11~_{-0.09}^{+0.08}$
	& ...
	& ...
	& \hspace{0.6em}$0.77~_{-0.07}^{+0.07}$
	& \hspace{0.6em}$0.59~_{-0.12}^{+0.14}$ \\ 
\hspace{0.5em}8.15 -- 13.45 
	& $-1.27~_{-0.04}^{+0.05}$
	& $-3.29~_{-6.71}^{+0.97}$
	& $106.80~_{-11.9}^{+12.5}$
	& \hspace{0.6em}$5.03~_{-0.15}^{+0.15}$
	&  $14.12~_{-0.67}^{+0.72}$ \\ 
 13.45 -- 18.65 
	& $-0.89~_{-0.02}^{+0.02}$
	& $-2.35~_{-0.08}^{+0.06}$
	& $87.03~_{-3.0}^{+3.2}$
	&  $24.12~_{-0.26}^{+0.24}$
	&  $95.69~_{-1.27}^{+1.28}$ \\ 
 18.65 -- 23.85 
	& $-1.27~_{-0.04}^{+0.04}$
	& $-2.70~_{-0.20}^{+0.12}$
	& $41.35~_{-2.0}^{+2.2}$
	&  $17.57~_{-0.23}^{+0.25}$
	&  $22.66~_{-0.73}^{+0.69}$ \\ 
 23.85 -- 29.15 
	& $-1.15~_{-0.03}^{+0.03}$
	& $-2.32~_{-0.06}^{+0.04}$
	& $52.45~_{-2.0}^{+2.4}$
	&  $33.34~_{-0.30}^{+0.31}$
	&  $66.74~_{-1.04}^{+1.04}$ \\ 
 29.15 -- 34.35 
	& $-1.67~_{-0.03}^{+0.02}$
	& $-4.57~_{-5.43}^{+1.86}$
	& $34.45~_{-1.8}^{+2.0}$
	&  $14.39~_{-0.24}^{+0.24}$
	&  $12.29~_{-0.23}^{+0.83}$ \\ 
 34.35 -- 39.65 
	& $-1.90~_{-0.06}^{+0.05}$
	& $-2.57~_{-7.43}^{+0.34}$
	& $19.81~_{-7.6}^{+5.2}$
	& \hspace{0.6em}$6.38~_{-0.16}^{+0.19}$
	& \hspace{0.6em}$4.99~_{-0.62}^{+0.55}$ \\ 
 39.65 -- 44.65 
	& $-2.01~_{-0.04}^{+0.04}$
	& ...
	& ...
	& \hspace{0.6em}$3.47~_{-0.15}^{+0.15}$
	& \hspace{0.6em}$3.50~_{-0.29}^{+0.30}$ \\
 44.65 -- 49.65 
	& $-1.87~_{-0.05}^{+0.05}$
	& ...
	& ...
	& \hspace{0.6em}$1.95~_{-0.13}^{+0.12}$
	& \hspace{0.6em}$2.85~_{-0.31}^{+0.31}$ \\
 49.65 -- 54.65 
	& $-2.06~_{-0.08}^{+0.08}$
	& ...
	& ...
	& \hspace{0.6em}$1.50~_{-0.13}^{+0.13}$
	& \hspace{0.6em}$1.30~_{-0.23}^{+0.26}$ \\
 54.65 -- 59.65 
	& $-2.14~_{-0.12}^{+0.11}$
	& ...
	& ...
	& \hspace{0.6em}$1.17~_{-0.12}^{+0.13}$
	& \hspace{0.6em}$0.82~_{-0.22}^{+0.25}$ \\
 59.65 -- 64.65 
	& $-2.36~_{+0.16}^{-0.14}$
	& ...
	& ...
	& \hspace{0.6em}$0.93~_{-0.12}^{+0.12}$
	& \hspace{0.6em}$0.35~_{-0.12}^{+0.16}$ \\
\enddata
\vskip -18pt
\tablecomments{Fluxes are in units of $10^{-7}$~erg/cm$^2$/s.
$^a \alpha$ is the Band $\alpha$ parameter for those time regions
where a Band $\beta$ value is specified; otherwise, it is the index for
a simple power-law fit.
\wxmonly~Only WXM data were used in the fit.
All errors are 90\%.
\label{table:specparams}
}
\end{deluxetable}